\documentclass[12pt]{iopart}

\usepackage{iopams}
\usepackage{setstack}
\usepackage[english]{babel}
\usepackage[latin1]{inputenc}
\usepackage[T1]{fontenc}
\usepackage{lmodern}
\usepackage{amssymb}
\usepackage{graphicx}                   
\usepackage{braket}
\usepackage{graphicx}
\usepackage{color}
\usepackage{bm}
\usepackage{latexsym}

\usepackage{epstopdf}

\newcommand{\A}{\,\hat{a}^{\dagger}}

\begin{document}

\title[Experimental observation of bulk and edge transport in photonic Lieb lattices]{Experimental observation of bulk and edge transport in photonic Lieb lattices}

\author{D. Guzm\'an-Silva$^{1}$, C. Mej\'ia-Cort\'es$^{1}$, M.A. Bandres$^{2}$,\\M.C. Rechtsman$^{2}$, S. Weimann$^{3}$, S. Nolte$^{3}$, M. Segev$^{2}$,\\A. Szameit$^{3}$, and R. A. Vicencio$^{1}$}

\address{$^1$Departamento de F\'{\i}sica, MSI-Nucleus on Advanced Optics, and Center for Optics and Photonics (CEFOP), Facultad de Ciencias, Universidad de Chile, Santiago, Chile}

\address{$^2$Physics Department and Solid State Institute, Technion, 32000 Haifa, Israel}

\address{$^3$Institute of Applied Physics, Friedrich-Schiller-Universit\"at Jena, Max-Wien-Platz 1, 07743 Jena, Germany}

\ead{vicenciorodrigo1@gmail.com}

\begin{abstract}
We analyze the transport of light in the bulk and at the edge of photonic Lieb lattices, whose unique feature is the existence of a flat band
representing stationary states in the middle of the band structure that can form localized bulk states. We find that transport in bulk Lieb lattices
is significantly affected by the particular excitation site within the unit cell, due to overlap with the flat band states. Additionally, we
demonstrate the existence of new edge states in anisotropic Lieb lattices. These states arise due to a virtual defect at the lattice edges and are
not described by the standard tight-binding model.
\end{abstract}



\maketitle

\section{Introduction}
A `flat band' is a dispersionless energy band composed of entirely degenerate states. As a result of this degneracy, any superposition of these
states (even a highly localized one) is completely static, displaying no evolution dynamics whatsoever. Flat bands are manifested in a variety of
condensed matter contexts, including ferromagnetic ground states of Hubbard models~\cite{Tasaki2008}, frustrated hopping models~\cite{Bergmann2008},
in the context of localization in the presence of magnetic fields or spin-orbit coupling~\cite{Vidal1998}, and the fractional quantum Hall
effect~\cite{Neupert2011}. With the advent of the recent experimental observations of photonic topological insulators~\cite{Rechtsman2013,Hafezi},
topological flat band models are now ripe for study in the optical domain. Indeed in optics, flat bands have direct technological relevance due to
their high degeneracy, and thus high photonic density-of-states, leading to enhanced light-matter interaction~\cite{Wang2009}. Additionally, for
Kagome systems single peak solitons have shown to bifurcate from the flat band at zero power threshold~\cite{PRAR13}, and a non-diffractive image
transmission scheme have been suggested in the linear regime~\cite{JOPT13}, showing the chance for concrete applications emerging from the flat band
properties.

In the Lieb lattice - a line-centered square lattice - a flat band touches two linearly dispersing intersecting bands, i.e., the flat band intersects
a single Dirac point. In the low-energy regime, the dispersion is described by a quasi-relativistic equation for spin-1 fermions~\cite{Shen2010}. It
is manifested in, e.g., the high-temperature tetragonal phases of the so-called cuprates: layered, ceramic materials that exhibit the phenomenon of
high-temperature superconductivity~\cite{Bednorz1986}. In this particular system several studies were conducted only recently, experimentally probing
its response to external gauge fields~\cite{Goldman2011} to nematic ferromagnetism~\cite{Zhang2012} and nonlinear conical
diffraction~\cite{Leykam2012}. The flat-band zero-energy edge modes in this lattice exhibit various peculiar features~\cite{Nita2013, Leykam2013,
Flach2013} and, hence, deserve particular attention.

Here, we analyze transport in Lieb lattices theoretically and experimentally, in the bulk and at the edges. We demonstrate that whether or not
flat-band modes are excited strongly impacts the transport in the bulk of the lattice. Additionally, we demonstrate the presence of Tamm-like edge
states that arise at a van Hove singularity in the band structure~\cite{Plotnik2014}. Finally, we show that our photonic realization holds promise
for a thorough analysis of various features of the Lieb lattice in terms of transport at and far off the lattice edges.

\section{Bulk transport}

\begin{figure}[htb]
\centering
\includegraphics[width=\columnwidth]{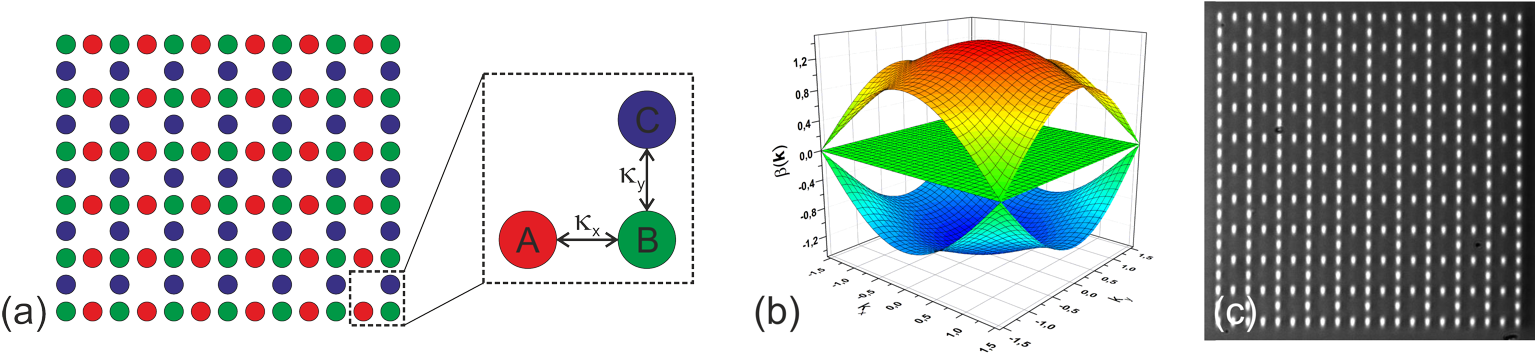}
\caption{(a) Sketch of a Lieb lattice with three elements per unit cell. The hopping between $A$ and $B$ is described by $\kappa_x$ (horizontal
direction), and between $B$ and $C$ by $\kappa_y$ (vertical direction). (b) The dispersion relation of the isotropic Lieb lattice
($\kappa_x=\kappa_y$) in the tight-binding approximation (Eq.\ref{tightbindingband}) shows three bands: a flat band at $\beta=0$ and two conically
intersecting ones, with all three bands intersecting at the M point ($k_x=k_y=\pi/2$). (c) Microscope image of a laser-written Lieb lattice in a
fused silica glass wafer.\label{fig:1}}
\end{figure}

The structure of the Lieb lattice is shown in Fig.~\ref{fig:1}(a), where it is evident that the unit cell consists of three sites $A$, $B$, and $C$.
As in honeycomb lattices (also known as ``photonic graphene") \cite{Peleg2007}, the equation describing the propagation of light in this system is
%
\begin{equation} \label{continuousmodel}
\hspace{-1.5cm} i \frac{\partial}{\partial z} \psi(x,y,z) = -\frac{1}{2k_0n_0}\nabla_{\bot}^2 \psi(x,y,z) - k_0\Delta n(x,y)\psi(x,y,z)\equiv
H_{\mathrm{cont}} \psi(x,y,z) \nonumber
\end{equation}
where $z$ is the longitudinal propagation distance into the photonic lattice (playing the role of time in the analogous Schr\"{o}dinger equation);
$\psi$ is the envelope of the electric field, defined by $E(x,y,z)=\psi(x,y,z)e^{i(k_0 z-\omega t)}$ ($E$ is the electric field, $k_0$ the wavenumber
in free space, $\omega=ck_0/n_0$ and $c$ is the speed of light in vacuum); $\Delta n(x,y)$ is the refractive index structure defining the photonic
lattice, while $n_0$ is the refractive index of the material in which the lattice is embedded; $\nabla_{\bot}^2=\partial_x^2+\partial_y^2$ is the
transverse Laplacian operator; $H_{\mathrm{cont}}$ as defined in equation (1) corresponds to the continuum Hamiltonian for wave propagation in a
photonic lattice. When the refractive index structure is composed of highly confined waveguides, each with a single bound state, we employ a
tight-binding approximation. Since within a given unit cell hopping takes place only between $A$-$B$ and $B$-$C$ nearest neighbors, the corresponding
tight-binding Hamiltonian reads
\begin{equation} \label{tightbindingmodel}
  H_{\mathrm{TB}} =  \sum_{\mathbf{R}_n, \mathbf{\delta}_j} \left ( \kappa_x a^{\dagger}_n b_n + \kappa_y b^{\dagger}_n c_n + h.c. \right ) \; ,
\end{equation}
with $a^{\dagger}_n$, $b^{\dagger}_n$, $c^{\dagger}_n$ as the creation operators in the $n$-th unit cell on the $A$, $B$ and $C$ sites, respectively.
The summation takes place over $\mathbf{R}_n$ (the position of the unit cell), and $\mathbf{\delta}_j$ (the vectors connecting the neighbors via the
inter-site coupling $\kappa_{x,y}$ in the $x$- and $y$-direction, respectively). In Fourier space, this becomes
\begin{equation}
  \mathcal{H}_{\mathrm{TB}} = 2 \left ( \begin{array}{c c c}
    0 & \kappa_x\cos k_x & 0\\
    \kappa_x\cos k_x & 0 & \kappa_y\cos k_y\\
    0 & \kappa_y\cos k_y & 0
  \end{array} \right )
\end{equation}
with $k_{x,y}\in \lbrack -\pi/2, \pi/2\rbrack$ (first Brillouin zone). From this, one readily obtains the dispersion relation, i.e., the spectrum
\begin{equation}\label{tightbindingband}
  \beta(\mathbf{k}) = 0; \pm 2 \sqrt{\kappa^2_x \cos^2 k_x + \kappa^2_y \cos^2 k_y} \; .
\end{equation}
%
The full spectrum (shown in Fig. \ref{fig:1}(b) for $\kappa_x = \kappa_y = 1$) consists of three bands: a flat band ($\beta(\mathbf{k}) = 0$) and two
conically intersecting bands, with all three bands intersecting at the M point ($k_x=k_y=\pi/2$). Note that for each eigenstate at Bloch wavevector
${\bf k}$ in the top band with eigenvalue $\beta$, there exists an eigenstate also at ${\bf k}$ with eigenvalue $-\beta$; this is known as
particle-hole symmetry~\cite{He2013}. The eigenfunctions of the finite Lieb lattice with periodic and vanishing boundary conditions are given in Ref.~\cite{Nita2013}. Importantly, any superposition of eigenstates belonging to the flat band is also an eigenstate of the system, meaning that it is
static and does not contribute to transport. Equivalently, the group velocity in the flat band is zero at all Bloch wave vectors, therefore any
wavepacket is necessarily diffraction-free. Consequently, to have transport, states in the other bands have to be excited. We also note that the flat
band is not destroyed by any anisotropy $\kappa_x \not= \kappa_y$, being an intrinsic property of this lattice in the nearest-neighbor coupling
limit~\cite{Leykam2012}.

In general, linear modes of any periodic structure are completely extended. However, flat band systems allow the formation of very localized
eigenmodes as a destructive linear combination of extended linear wavefunctions~\cite{Berg2008}. In the case of the Lieb lattice, in any square of
the lattice (formed by four $B$ sites at the corners, two $C$ sites at the upper and lower sides and two $A$ sites at the right and left sides of the
square) a stationary localized mode can be created by taking the four $B$ sites  strictly zero, the two $C$ sites with an amplitude $\psi_C$, while
the last two $A$ with an amplitude $\psi_A=-\kappa_y \psi_C /\kappa_x$ and the rest of lattice sites are zero.

For our experiments, we fabricate Lieb lattices employing the laser direct-write technology~\cite{Szameit2010a}. To this end, ultrashort laser pulses
(Coherent Inc. Mira/RegA, $\lambda=800$nm, $t_{p}\approx 200$fs at 100kHz repetition rate) are tightly focused through a microscope objective
(25$\times$, NA = 0.35) into a 10cm long transparent fused silica wafer and translated by means of a high-precision positioning system (Aerotech
Inc.). A microscope front view of such a lattice, which consists of $341$ waveguides with nearest-neighbor spacing of 20$\mu$m, is shown in
Fig.~\ref{fig:1}(c). Note that due to the slightly elliptical form of our waveguides, we find that the vertical inter-site coupling is slightly
stronger than the horizontal one, i.e. the lattice is effectively uniaxially strained and anisotropic with $\kappa_y \approx 1.5 \kappa_x$. In our
experiments, laser light at $\lambda = 637$ nm was launched into a single waveguide at the input facet of the array using a standard microscope
objective. At the output facet of the sample, intensity patterns were recorded with a CCD camera.

\begin{figure}[htb]
\centering
\includegraphics[width=\columnwidth]{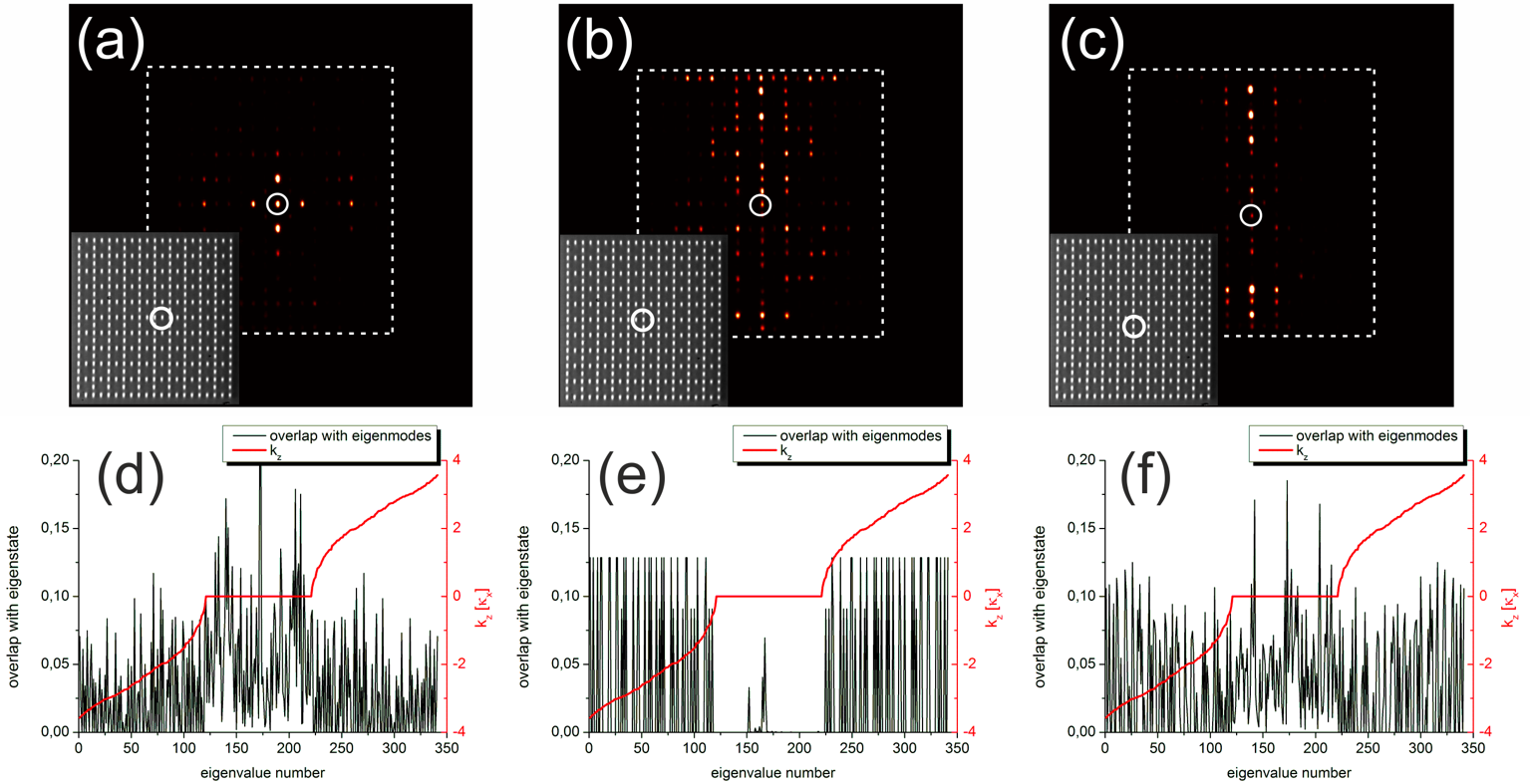}
\caption{Experimentally observed light distribution at the output facet of the lattice, when exciting (a) the $A$-site, (b) the $B$-site, and (c) the
$C$-site. The overlap with the eigenmodes in the three bands is shown in (d) for the $A$-site, (e) for the $B$-site, and (f) for the $C$-site. It is
evident that transport is stronger when the flat band is least populated at the initial plane of excitation. \label{fig:2}}
\end{figure}
We begin by launching light into the bulk of the lattice, i.e., far away from the edges. As there are three elements per unit cell (see
Fig.~\ref{fig:1}a), we study the propagation arising from excitations in each of the three positions $A$, $B$, and $C$. The corresponding diffraction
patterns obtained in our experiments are shown in Figs.~\ref{fig:2}(a),(c),(e). One can clearly see that the impact of the excitation location on the
diffraction pattern is significant, depending on where in the unit cell the lattice is excited. The reason lies in the unique structure of the Lieb
lattice, where a single site excitation (that corresponds to a unit impulse in real space) does not excite all bands - in contrast to other lattices,
like square, hexagonal or honeycomb lattices. To illustrate this effect, we plot the projections $P=\langle k,j | \psi^{excit}\rangle$ of the impulse
excitation $|\psi^{\mathrm{excit}}\rangle$ at the $A$, $B$, or $C$ sites on the eigenfunctions $|k,j\rangle$ of the three bands ($j$ is the band
index) in Figs.~\ref{fig:2}(b),(d),(f), and compare them to the experimental data. It is evident that, when exciting the $B$ site
[Fig.~\ref{fig:2}(e)], the projections on the flat band are minimal (equal to $1/(N+1)^2$, where $N$ is the number of complete unit cells on one side
of a finite square Lieb lattice, i.e., $N=10$ in our case) and, hence, transport is strong. This is clearly observed in the experiment
[see~\ref{fig:2}(b)], where the light spreads across many lattice sites following the 10~cm propagation in the lattice. However, when exciting the
$A$ site the situation changes drastically [Fig. \ref{fig:2}(d)]. In this case, the projections on the flat band dominate and, therefore, the
transport is minimal. The experiment shows exactly this: the spreading is very weak, and the light covers only a few lattice sites after propagating
through the lattice [see Fig.~\ref{fig:2}(a)]. Intermediate spreading is observed when launching light into lattice site $C$, as shown in
Fig.~\ref{fig:2}(c), as a consequence of the partial overlap with the flat band [see Fig.~\ref{fig:2}(f)].

\begin{figure}[htb]
\centering
\includegraphics[width=12cm]{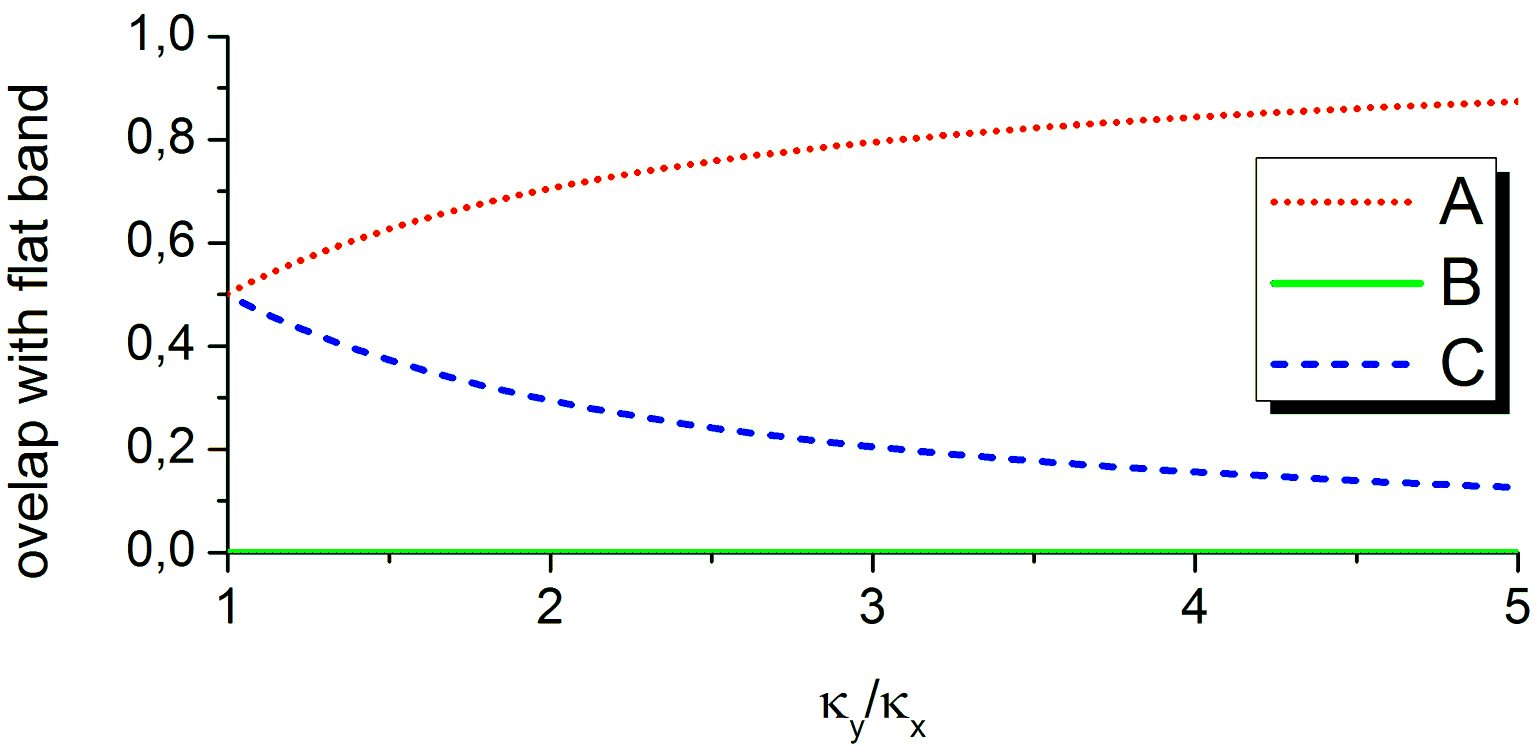}
\caption{The overlap of single-site excitation with flat band modes, for excitation of the $A$-site (red dotted line), $B$-site (green solid line),
and $C$-site (blue dashed line). \label{fig:3}}
\end{figure}

At this point it is interesting to examine the transport through the Lieb lattice when strain is applied. Specifically, it has been shown that
inhomogeneous strain applied to honeycomb lattices can give rise to Landau level splitting \cite{Rechtsman2013b}, robust Klein tunneling
\cite{Bahat-Treidel2010}, topological transitions \cite{Rechtsman2013c}, and other unexpected phenomena. It is therefore instructive to look at the
projections on the flat band as a function of the strain when exciting either the $A$, $B$, or $C$ sites in the infinite Lieb lattice. Our
calculations are summarized in Fig.~\ref{fig:3}. Interestingly, when exciting a $B$ site, the population of the flat band always vanishes, for all
values of the anisotropy. Localized flat band modes have $B$ sites with zero amplitude, therefore launching light into this site will always result
in strong transport, irrespective of the ratio $\kappa_y/\kappa_x$. However, the situation is different for an excitation of the $A$ and $C$ sites.
Whereas for the isotropic case ($\kappa_x =\kappa_y$), the overlap with the flat band for both sites is exactly the same (due to symmetry reasons),
for the anisotropic case ($\kappa_y > \kappa_x$) the flat band is more populated for an $A$-site excitation (resulting in decreasing transport) and
less populated for a $C$-site excitation (yielding enhanced transport). (Anisotropic $\beta=0$ modes have $|\psi_A|>|\psi_C|$, being therefore easier
to excite them in sites $A$).

\section{Edge transport}

\begin{figure}[htb]
\centering
\includegraphics[width=\columnwidth]{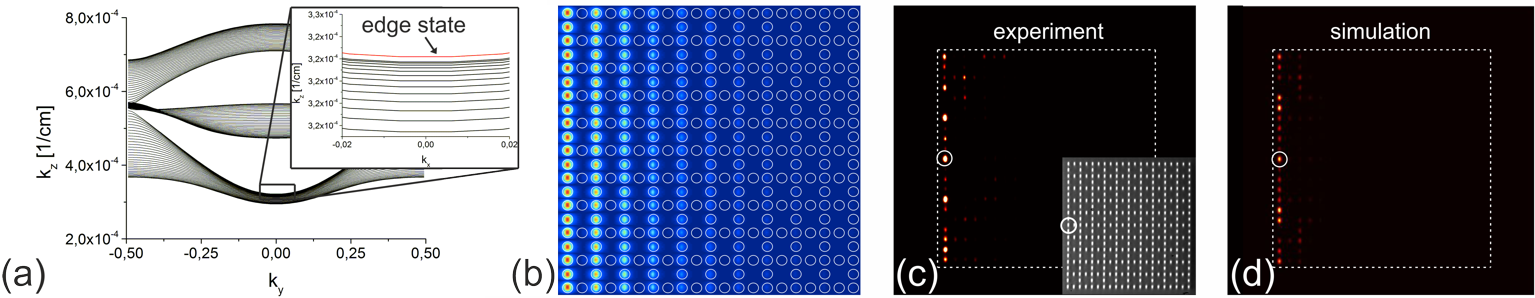}
\caption{(a) Edge band structure of a strained Lieb lattice ($\kappa_y = 1.5\kappa_x$) using the continuous model Eq.~\ref{continuousmodel}, showing
the appearance of edge states (red line). (b) The intensity distribution of the edge state in the Lieb lattice. (c) Experimental observation of the
edge states after an impulse excitation. (d) Simulation of the light diffraction at the edge. \label{fig:4}}
\end{figure}

Notwithstanding the bulk transport, we pay particular attention to the diffraction of light at the edge of the lattice. It is important to note that,
within tight-binding model (Eq.~\ref{tightbindingmodel}) there are no edge states present, in the isotropic as well as in the anisotropic case.
However, we surprisingly find that edge states arise in anisotropic Lieb lattices in the continuous model (Eq.~\ref{continuousmodel}). In
Fig.~\ref{fig:4}(a) we present a plot of the band structure at the edge of the Lieb lattice by diagonalizing $H_{cont}$ as given by
Eq.~\ref{continuousmodel}, showing clearly the existence of states residing in the band gap that are localized at the lattice edge. The underlying
mechanism for the formation of these edge states is similar to the one recently reported for ``photonic graphene''~\cite{Plotnik2014}. Note that
these calculations are performed by full diagonalizing $H_{cont}$, as given in Eq.~\ref{continuousmodel}, as opposed to simply examining the spectrum
of the tight-binding model.

To explain these edge states in the Lieb lattice, one has to take into account the fact that the effective on-site propagation constant of each mode
is modified by its neighbouring sites. This modification differentiates between the $B$-sites in the bulk, where they have four neighbours, and the
edge, where they have only three. Although this difference is small and is usually neglected in tight-binding theories, it comes into play where the
modes are highly degenerate at the van Hove singularity. At the degeneracy point, any slight edge perturbation takes the edge mode out of the band
and creates a Tamm-like edge state as shown in Fig.~\ref{fig:4}(b). We emphasize again that this state does not result from any real surface
perturbation but rather from the specific surface structure along the edge. We classify this state as a ``Tamm state''~\cite{Tamm1932} as opposed to
a ``Shockley state''~\cite{Shockley1939}, because it does not arise due to a band crossing, the criterion for the emergence of Shockley states. It is
important to note that these states do not have a topological origin, and only arise from the `effective defect' at the edge. Although Tamm states
are conventionally associated with surface perturbations or defects (that are inherent in the system, or that arise from
modulation~\cite{Szameit2008}), in the present case no defects whatsoever are present. Instead, what happens here is that the edge itself acts as a
sufficiently strong defect to localize light on the edge. This effect can not be accounted for in the standard tight-binding model and appears only in the
continuum model. The experimental observation of strong edge diffraction in the anisotropic Lieb lattice is presented in Fig.~\ref{fig:4}(c), the
simulations using Eq.~\ref{continuousmodel} are shown in Fig.~\ref{fig:4}(d).

\section{Conclusions}
In conclusion, we presented a theoretical and experimental study of transport of light in photonic Lieb lattices. We find that the transport in the
bulk is significantly affected by which site in the unit cell is excited, due to different overlap with the flat band (since any superposition of
flat band states does not diffract and can form localized bulk states). Importantly, an excitation of the $B$-site in the unit cell leaves the flat
band essentially empty, resulting in maximal transport. We also show that the population of the different bands is a function of the anisotropy of
the lattice. Moreover, we demonstrate the existence of new edge states in anisotropic Lieb lattices that arise due to a virtual defect at the lattice
edges and are not manifested in the standard tight-binding model. Our findings experimentally prove that the Lieb lattice exhibits various peculiar
and unique features concerning wave transport, which might find their way into new concepts for imaging, routing and switching.

\section*{Acknowledgements}
The authors wish to thank the German Ministry of Education and Research (Center for Innovation Competence programme, grant 03Z1HN31), and the
Thuringian Ministry for Education, Science and Culture (Research group Spacetime, grant 11027-514), FONDECYT Grant 1110142, Programa ICM P10-030-F,
and Programa de Financiamiento Basal de CONICYT (FB0824/2008).


\end{document}